\documentclass[aps,pra,showpacs,groupedaddress,superscriptaddress,twocolumn]{revtex4}
\usepackage{amsfonts}
\usepackage{graphicx}
\usepackage{latexsym}
\usepackage{makeidx} 
\usepackage{amsmath} 
\usepackage{amssymb}
\usepackage{graphicx}
\usepackage{color}
\begin{document}

\title{Eigenvalue decomposition method for photon statistics of frequency filtered fields \\
 and its applications to quantum dot emitters}
\author{Kenji Kamide}
\email{kamide@iis.u-tokyo.ac.jp}
\affiliation{%
 Institute for Nano Quantum Information Electronics (NanoQuine), University of Tokyo, Tokyo 153-8505, Japan}
\author{Satoshi Iwamoto}
\author{Yasuhiko Arakawa}%
\affiliation{%
 Institute for Nano Quantum Information Electronics (NanoQuine), University of Tokyo, Tokyo 153-8505, Japan}
\affiliation{%
 Institute of Industrial Science, University of Tokyo, Tokyo 153-8505, Japan
}%
\date{\today}
\begin{abstract}
A simple calculation method for photon statistics of frequency-filtered fields is proposed. This method, based on eigenvalue decompositions of superoperators, allows us to study effects on the photon statistics of spectral filtering by various types of filters, such as Gaussian and rectangular filters as well as Lorentzian filters, which is not possible by conventional approaches. 
As an example, this method is applied to a simulation of quantum dot single-photon emitters, where we found the efficient choice of the filter types to have pure single photons depends on the excitation conditions i.e. incoherent or coherent (and resonant) excitations.
\end{abstract} 
\maketitle
\section{Introduction}

High quality single photon (SP) sources, which emit one photon at a time with high purity and high rate, are essential for the realistic and reliable application to quantum information science and technologies~\cite{BB84, Waks}.
As efficient solid-state SP sources, semiconductor quantum dots (QDs) are promising candidate systems in solids and have been attracting attention for a number of advantages: the well-defined quantized states, high controllability in the emission wavelength, high brightness even enhanced by embedding them in nanocavities, the emission-site controllability, and possible current injection operations~\cite{Buckley, Santori, Takemoto, Kako, Holmes, Nakaoka, Englund, Strauf, Muller, Ates}. 

\begin{figure}[b]
\begin{center}  
\includegraphics[scale=0.65]{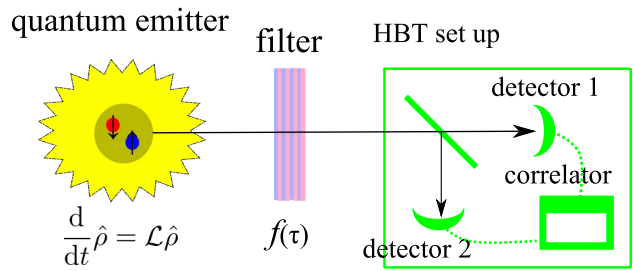} {} 
\end{center} 
\vspace{-2 mm}
\caption{\label{fig1} Quantum emitter and detection system. The emission dynamics of the quantum emitter is given by the quantum master equation, $\frac{{\rm d}}{{\rm d}t}\hat{\rho}=\mathcal{L}\hat{\rho}$. The effect of spectral filter is described by the correlation function $f(\tau)$, and the photons passed through the filter enter into the HBT setup~\cite{HBT} for the $g^{(2)}$ correlation measurement.}
\end{figure}

However, in QD SP emitters, a number of emission lines are typically present due to the multiple transition levels and also to other QDs in a sample, degrading the SP purity. To avoid the degradation, spectral filters (Fig. \ref{fig1}) are usually used for selecting the relevant emission e.g. an exciton emission, and filter out the spectrally separated irrelevant emissions, such as, the biexciton-exciton emission~\cite{Kamide} and charged excitons.
The role of using a spectral filter is to prevent the detection of irrelevant emissions spectrally separated in the frequency domain.
At the same time, due to the frequency-time uncertainty, using narrow spectral filters inevitably widens the detection field in the time domain, leading to degradation in the time resolution and also in the purity of the SP emissions.
In this way, spectral filter modifies the filtered field both in the frequency and time domains, and therefore, the filtering effect on the photon statistics  (which basically is a multiple-time correlation function given in the time domain) is not so simply understood, especially for quantum emitters.

Theoretical and experimental studies of filtering effects on the photon statistics have recently been attracting attention~\cite{Valle, Valle2, Gonzalez-Tudela, Munoz, Gonzalez-Tudela2}. 
These studies were triggered by the development in the theoretical treatment, a versatile calculation method proposed by E. del Valle {\it et al.}~\cite{Valle}.
In this method, the spectral filtering process is effectively replaced with the inclusion of probe systems coupling weakly to the system.
A great advantage of this method over the former theory~\cite{Cresser, Nienhuis, Nienhuis2} is that the complication in calculation coming from the time orderings of operators can be avoided, allowing the calculation of higher-order correlation functions ($n \ge 3$).
However, in this method, the type of spectral filters the method can treat is restricted to Lorentzian filters, since the spectral filter is mimicked by a Lorentzian density of states of a probe system under the Markov decay process.
The former analytic, but approximate, approach~\cite{Nienhuis} also treats only Lorentzian filters due to its simplicity.
Therefore, the effect of spectral filtering on the photon statistics has been investigated only for Lorentzian filters so far.

In this paper, we propose a simple calculation method that allows for theoretical treatment of a variety of spectral filters in order to deepen the understanding of filter effect on the photon statistics.
In Sec.~\ref{secII}, we introduce the calculation method based on superoperator eigenvalue decomposition, and find the exact expressions for the second-order correlation function for Gaussian and rectangular filters as well as the Lorentzian filters. 
As far as we know, this is the first time to apply the superoperator eigenvalue decomposition technique to the higher-order correlation functions, whereas it has been applied to the calculation of the first-order correlation functions e.g. in a calculation of the M{\"o}ssbauer spectra~\cite{Clauser}.
While our method allows for the treatment of types of filters, it directly treats the operator ordering problem, and thus the difficulty in calculating high order correlation functions is not removed. In this sense, our method is complementary to the previous theory~\cite{Valle}. 
In Sec.~\ref{sec: applications}, as an example, we show a numerical simulation applied to QD SP emitter systems, where we found the efficient choice of the filter types for purifying the single photons depends on the excitation conditions i.e. incoherent or coherent (and resonant) excitations.

We note that the effect of the background noise which is not related to the system dynamics is out of the scope in the theoretical framework. Throughout the paper, we set $\hbar =1$ for simplicity unless otherwise specified.

\section{\label{secII}Superoperator eigenvalue decomposition method for photon statistics of frequency filtered fields}

\subsection{Definition of the problem}
Here, we will define the problems to solve.
The system we consider consists of an emitter, a spectral filter, and detection system as shown in Fig.~\ref{fig1}.
Photons emitted from the quantum emitter are detected by Hanbury-Brown Twiss (HBT) setups~\cite{HBT} for the second-order intensity correlation measurement, and before the detection, the photons passed through a spectral filter whose response is described by the filter correlation function $f(\tau)$. Alternatively, the detection can be performed with high-speed streak camera with high time resolution (less than a few picoseconds), which is now becoming a powerful detection system for the study of photo-counting statistics~\cite{Wiersig}.
Throughout the paper, we assume the effect of back reflection at the filter surface on the emitter system can be neglected~\cite{Cresser}.
In this case, the dynamics of the quantum emitter and the emission field is given by the Hamiltonian, $\hat H$, for the emitter, Lindblad-type superoperators, $\mathcal{L}_{\eta}$, for decay and pump processes labeled by $\eta$ with the rates, $\gamma_\eta$, and the resulting quantum master equation~\cite{Carmichael},
\begin{eqnarray}
\frac{{\rm d}}{{\rm d}t} \hat{\rho} 
=i[\hat{\rho} ,\hat{H} ] +\sum_\eta \gamma_{\eta} \mathcal{L}_{\eta}\hat{\rho}
\equiv \mathcal{L}\hat{\rho}. \label{eq:QME}
\end{eqnarray}
The emission field operator, $\hat{E}^{\pm}$, at the exact emission time, $t$, is given by the Heisenberg operator (accounting for the system dynamics except for the filter and detection systems), $\hat{E}^{\pm}(t)$, and the frequency-filtered field $\hat{E}^{\pm}_F(t)$ to be detected at a time $t$ is given by
\begin{eqnarray}
\hat{E}^{-}_F(t)=\int_0^\infty f(\tau)\hat{E}^{-}(t-\tau) {\rm d} \tau, \nonumber \\
\hat{E}^{+}_F(t)=\int_0^\infty f^\ast(\tau)\hat{E}^{+}(t-\tau) {\rm d} \tau, 
 \label{eq:FilteredField} 
\end{eqnarray} 
\begin{figure}[tb]
\begin{center}  
\includegraphics[scale=0.3]{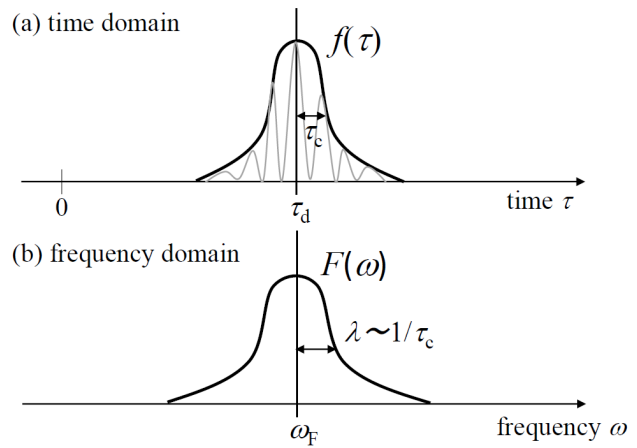} {} 
\end{center} 
\vspace{-7 mm}
\caption{\label{fig2} Schematics of filter function: (a) $f(\tau)$ in time domain and (b) $F(\omega)$ in frequency domain.}
\end{figure} 
The time region of the integration is physically restricted to $\tau>0$ by the causality, and the correlation function in the time domain, $f(\tau)$, has the peak at a delay time, $\tau=\tau_d$, corresponding to the optical path length between the emitter and the filter, and it has a width $\tau_c$ corresponding to the filter correlation time (Fig.~\ref{fig2} (a)). The filter function $F(\omega)$ in the frequency domain, is centered at $\omega_F$ and has a bandwidth $\lambda$ roughly equal to the inverse of the correlation time, $\lambda \sim 1/\tau_c$ (Fig.~\ref{fig2} (b)). 
The equations (\ref{eq:QME}) and (\ref{eq:FilteredField}) are the most general expression for system dynamics and the filtered emission field, and thus, can be directly applied to any emitters and any filters. 
For example, in case of a resonantly driven two-level atom (transition energy $\omega_{\rm A}$, Rabi frequency $\Omega_{\rm R}$, and the laser frequency $\omega_L$), $\hat{H}=\omega_{\rm A} \hat{\sigma}^+ \hat{\sigma}^- + \Omega_{\rm R} e^{-i\omega_L t} \hat{\sigma}^+ + \Omega_{\rm R}^\ast e^{i\omega_L t} \hat{\sigma}^-$, the spontaneous emission decay is included with the rate $\gamma_\eta=\gamma_{sp}$ and superoperator $\mathcal{L}_{\eta}=\mathcal{L}_{\hat{\sigma}^-}$ in standard notation, and the emission field is $\hat{E}^{\pm}(t)=\hat{\sigma}^{\pm}(t)$.

The $n$-th order normalized intensity correlation function to be evaluated is then given by
\begin{eqnarray}
&&g^{(n)}(t_1,t_2, \cdots , t_n) \nonumber \\
&& \quad =\frac{\langle \mathcal{T}_+ \mathcal{T}_-
\hat{E}^{+}_F(t_1)   \cdots \hat{E}^{+}_F(t_n)  
\hat{E}^{-}_F(t_n)  \cdots \hat{E}^{-}_F(t_1) \rangle}
{\prod_{j=1}^{n}
 \langle \hat{E}^{+}_F(t_j)  \hat{E}^{-}_F(t_j)  \rangle}, \nonumber  \label{eq:Gn-general} \\
\end{eqnarray} 
where $\mathcal{T}_-$ and $\mathcal{T}_+$ are time ordering and anti-ordering superoperators working on the Heisenberg annihilation and creation operators, respectively. 
The brackets mean the ensemble statistical average over the emitter states, and mathematically given by taking the trace after multiplying by the density matrix of the emitter system, $\langle \hat{O}  \rangle={\rm Tr} ( \hat{O} {\hat \rho} )$.  
In addition, from Eq.~(\ref{eq:FilteredField}), it is necessary in calculating Eq.~(\ref{eq:Gn-general}) to evaluate the operator products with different time arguments by using quantum regression theorem~\cite{Carmichael}.
The aim of this paper is to give a simple calculation method for the correlation function in Eq.~(\ref{eq:Gn-general}).

The time ordering operation has to be taken into account in Eq.~(\ref{eq:Gn-general}) when the effect of back reflection by the filter is negligible as we assumed here~\cite{Cresser}. 
However, the operation makes the calculation of the $n$-th order correlation function of large $n$ (like $n \ge 3$) complicated.
For $g^{(n)}(t_1=\cdots=t_n)$, the number of time arguments ($\tau_j$ with $j=1,\cdots n$ for ${\hat E}^+$ fields and $j=n+1,\cdots 2n$ for ${\hat E}^-$ fields) are $2n$. 
The number of different time orderings in the integration is reduced to $(2n)!/(n!)^2$ by a symmetry argument; The number is reduced from $(2n)!$ to $(2n)!/(n!)^2$ since the time ordering operators $\mathcal{T}_+$ and $\mathcal{T}_-$ sort the product of ${\hat E}^+$ fields and ${\hat E}^-$ fields of $n!$ different orderings, respectively, into one exclusive ordering. 
Therefore, the number of terms with different time orderings for the $n$-th order correlation function amounts to six for $n=2$, twenty for $n=3$, and seventy for $n=4$~\cite{Nienhuis}.

In our approach given below, we will finally obtain the analytic expression for the correlation function in Eq.~(\ref{eq:Gn-general}), whereas the time ordering process is directly treated, hence the difficulty is not removed.
Therefore, the higher order correlation function with $n \ge 4$ is too computationally expensive. 
In this sense, this method is limited to the application to the correlation functions with $n \le 3$ in realistic calculation, while the recently proposed method~\cite{Valle} can avoid the complicated time ordering operation to be able to simulate photon statistics to the higher order.

However, as mentioned in the introduction, this method allows us to have analytic results for general types of the filter function $f(\tau)$, whereas the previous method~\cite{Valle} can treat only the Lorentzian filters. 
In this sense, our method is complementary to the other methods~\cite{Nienhuis, Valle}, and this makes possible the comparison of the efficiency in optimizing the photon statistics for different types of filters.

\subsection{Superoperator eigenvalue decomposition method}
Here, we will introduce our method based on superoperator eigenvalue decompositions.
As an example for $n=2$, we will obtain a general expression for $g^{(2)}(\tau=t_1-t_2=0)$.

{\it Superoperator eigenvalue decomposition}---
According to the quantum regression formula~\cite{Carmichael}, different-time correlation functions can be calculated by the same equation as the density matrix equation in Eq.~(\ref{eq:QME}). Thus, any time dependent operator ${\hat O} (t)$ satisfies $\frac{d}{dt} {\hat O}=\mathcal{L} {\hat O}$. 
The matrix equation can be written in a linear equation $\frac{d}{dt} \vec{O}={\bf L} \vec{O} $ after reforming the operator ${\hat O}$ into a vector form $\vec{O}=({\hat O}_{1,1}, \cdots ,{\hat O}_{1 ,N_c},{\hat O}_{2,1}, \cdots ,{\hat O}_{2,N_c}, \cdots ,{\hat O}_{N_c,1}, \cdots ,{\hat O}_{N_c,N_c})$. 
The length of $\vec{O}$ is $N_c^2$, and the Liouvillian matrix ${\bf L}$ (originally the superoperaor $\mathcal{L}$) has dimension, $N_c^2 \times N_c^2$. 
Therefore, ${\bf L}$ has $N_c^2$ eigenvalues, $\Omega$, which are in general complex values with ${\rm Re}(\Omega) \le 0$.
The corresponding right and left eigenvectors, $\vec{v}_\Omega$ and $\vec{u}_\Omega^T$, are defined here as
\begin{eqnarray}
{\bf L} \vec{v}_\Omega=\Omega \vec{v}_\Omega, \quad 
\vec{u}_\Omega^T  {\bf L} =\vec{u}_\Omega^T \Omega.
\end{eqnarray} 
Therefore, if the eigenvalues are non-degenerate, the operator in vector form, $\vec{O}$, is decomposed into the eigenvectors:
\begin{eqnarray}
\vec{O}&=&\sum_{\Omega} C(\Omega)  \vec{v}_\Omega, 
\end{eqnarray}
where $C(\Omega) =(\vec{u}_\Omega^T \cdot \vec{O} )/ (\vec{u}_\Omega^T \cdot \vec{v}_\Omega )$.
From the above expression, we obtain the spectral decomposition of the vector by $\vec{O}=\sum_{\Omega} [\vec{O}](\Omega)$ with $[\vec{O}](\Omega) \equiv C(\Omega)  \vec{v}_\Omega$, whose matrix form is written as $\hat{O}=\sum_{\Omega} [\hat{O}](\Omega)$.  
The merit of using the eigenvalue decomposition is that the time evolution of the operators are explicitly given by
\begin{eqnarray}
\hat{O} (t) \equiv e^{\mathcal{L} t}[\hat{O}]=\sum_{\Omega} [\hat{O}](\Omega) e^{\Omega t}. \label{eq:decomposition}
\end{eqnarray} 

{\it Average filtered-field intensity}---
Now, we apply the eigenvalue decomposition technique to the filtered-field intensity, $\langle \hat{E}^{+}_F(t_j)  \hat{E}^{-}_F(t_j)  \rangle ={\rm Tr} \left(\hat{E}^{+}_F(t_j)  \hat{E}^{-}_F(t_j) {\hat \rho}_{SS} \right) $ in Eq.~(\ref{eq:Gn-general}), whereas the system is assumed to be in the steady state ${\hat \rho}(t_j)={\hat \rho}_{SS}$. 
Inserting Eq.~(\ref{eq:FilteredField}), we have
\begin{eqnarray}
&&\langle \hat{E}^{+}_F(t_j)  \hat{E}^{-}_F(t_j)  \rangle=
 \int_0^\infty \int_0^\infty {\rm d}{\tau_1} {\rm d}{\tau_2} \ f^\ast (\tau_1) f (\tau_2) \nonumber \\
&& \qquad \times  
{\rm Tr} \left(\hat{E}^{+}(t_j-\tau_1)  \hat{E}^{-}(t_j-\tau_2) {\hat \rho}_{SS} \right) \nonumber \\
&=& \iint_{\tau_2>\tau_1>0} {\rm d}{\tau_1} {\rm d}{\tau_2} \ f^\ast (\tau_1) f (\tau_2)
{\rm Tr} \left(\hat{E}^{+}e^{\mathcal{L}\tau_{21} } [\hat{E}^{-} {\hat \rho}_{SS}] \right) \nonumber \\
&+& \iint_{\tau_1>\tau_2>0} {\rm d}{\tau_1} {\rm d}{\tau_2} \ f^\ast (\tau_1) f (\tau_2)
{\rm Tr} \left(\hat{E}^{-}e^{\mathcal{L}\tau_{12} } [{\hat \rho}_{SS}\hat{E}^{+}] \right), \nonumber \\ \label{eq:Intensity}
\end{eqnarray} 
with $\tau_{ij} \equiv \tau_i-\tau_j$. Applying Eq.~(\ref{eq:decomposition}), the filtered-field intensity is expressed in the form
\begin{eqnarray}
\langle \hat{E}^{+}_F(t_j)  \hat{E}^{-}_F(t_j)  \rangle
=\sum_{\Omega} s(\Omega)q(\Omega)+s^\ast(\Omega)q^\ast(\Omega), \label{eq:IntensityGen} 
\end{eqnarray}
where we noticed that the first and second terms in the r.h.s. of Eq.~(\ref{eq:Intensity}) are the conjugate pairs. The coefficients are given by
\begin{eqnarray}
s(\Omega)&=&\iint_{\tau_2>\tau_1>0} {\rm d}{\tau_1} {\rm d}{\tau_2} \ f^\ast (\tau_1) f (\tau_2) e^{\Omega \tau_{21} } ,  \label{eq:IntensityGenS} \\
q(\Omega)&=&{\rm Tr} \left(\hat{E}^{+} [\hat{E}^{-} {\hat \rho}_{SS}](\Omega) \right) .  \label{eq:IntensityGenQ}
\end{eqnarray}
This is the general form of the superoperator eigenvalue decomposition for the filtered-field intensity. 
If the filter bandwidth $\lambda$ is set small, the intensity $\langle \hat{E}^{+}_F(t_j)  \hat{E}^{-}_F(t_j)  \rangle$ as a function of the central frequency $\omega_F$ is the emission spectrum. 
This eigenvalue decomposition method was previously applied to the calculation of the M{\"o}ssbauer spectra~\cite{Clauser} as a first-order correlation function. 
The method shown here is essentially the same as that shown in the paper.
However, we will now apply this method to the second-order correlation function.

{\it Average filtered-field intensity correlation}---
Next, we compute the second-order correlation function at zero delay assuming the steady state, 
\begin{eqnarray}
&& \left. \langle \mathcal{T}_+ \mathcal{T}_-
\hat{E}^{+}_F(t_1)    \hat{E}^{+}_F(t_2)  
\hat{E}^{-}_F(t_2)  \hat{E}^{-}_F(t_1) \rangle \right|_{t_1=t_2}\nonumber  \\
&=& 2^2 \iiiint_{
\begin{subarray} '
\tau_1>\tau_2>0 \\
\tau_4>\tau_3>0
\end{subarray}}  {\rm d}{\bf \tau}^4    f^\ast (\tau_1) f^\ast (\tau_2) f (\tau_3)f (\tau_4)  \nonumber \\ 
& \times & {\rm Tr} \left( \hat{E}^{+}(-\tau_1)   \hat{E}^{+}(-\tau_2) 
\hat{E}^{-}(-\tau_3) \hat{E}^{-}(-\tau_4) \hat{\rho}_{SS} \right),  \label{eq:2ndCorrelation}
\end{eqnarray}
where ${\rm d}{\bf \tau}^4 \equiv  {\rm d}\tau_1  {\rm d}\tau_2 {\rm d}\tau_3 {\rm d}\tau_4$.
The region of the four-fold integration is divided into six regions with different time orderings: 
(i) $\tau_2<\tau_3<\tau_4<\tau_1$,   
(ii) $\tau_2<\tau_3<\tau_1<\tau_4$,   
(iii) $\tau_2<\tau_1<\tau_3<\tau_4$,   
(iv) $\tau_3<\tau_2<\tau_1<\tau_4$,   
(v) $\tau_3<\tau_2<\tau_4<\tau_1$,   
(vi) $\tau_3<\tau_4<\tau_2<\tau_1$.
Since the contributions from (i) and (iv), (ii) and (v), and (iii) and (vi), are complex conjugate pairs respectively, we have only to compute the integration over (i), (ii), and (iii). 

The second-order correlation function in Eq.~(\ref{eq:2ndCorrelation}),  ${\rm Tr} \left( \hat{E}^{+}(-\tau_1)   \hat{E}^{+}(-\tau_2) 
\hat{E}^{-}(-\tau_3) \hat{E}^{-}(-\tau_4) \hat{\rho}_{SS} \right) $, is expressed, by using quantum regression theorem and the superoperator eigenvlue decomposition, as
\begin{eqnarray}
&=& {\rm Tr} \left(  \hat{E}^{+} e^{\mathcal{L}\tau_{32} }
\left[ \hat{E}^{-}  e^{\mathcal{L}\tau_{43} }
\left[ \hat{E}^{-}  e^{\mathcal{L}\tau_{14} }
\left[ \hat{\rho}_{SS} \hat{E}^{+}  
\right]
\right]
\right] \right) \nonumber \\
&=&\sum_{\Omega_{1},\Omega_{2},\Omega_{3}} {\rm exp} (\Omega_{1} \tau_{14}+\Omega_{2}\tau_{43}+\Omega_{3} \tau_{32}) \label{eq:gen2ndCor1}
 \\
&&\times
 {\rm Tr} \left(  \hat{E}^{+}
\left[ \hat{E}^{-} 
\left[ \hat{E}^{-} 
\left[ \hat{\rho}_{SS} \hat{E}^{+}  
\right](\Omega_{1})
\right](\Omega_{2})
\right](\Omega_{3}) \right) \nonumber 
\end{eqnarray}
for (i) $\tau_2<\tau_3<\tau_4<\tau_1$. 
Similarly, it is 
\begin{eqnarray}
&=&\sum_{\Omega_{1},\Omega_{2},\Omega_{3}} {\rm exp} (\Omega_{1} \tau_{41}+\Omega_{2}\tau_{13}+\Omega_{3} \tau_{32}) \label{eq:gen2ndCor2}
\\
&&\times
 {\rm Tr} \left(  \hat{E}^{+}
\left[  \hat{E}^{-} 
\left[ \left[\hat{E}^{-}   \hat{\rho}_{SS} \right](\Omega_{1}) 
\hat{E}^{+}  \right](\Omega_{2})\right](\Omega_{3})
 \right) \nonumber 
\end{eqnarray}
for (ii) $\tau_2<\tau_3<\tau_1<\tau_4$, and 
\begin{eqnarray}
&=&\sum_{\Omega_{1},\Omega_{2},\Omega_{3}} {\rm exp} (\Omega_{1} \tau_{43}+\Omega_{2}\tau_{31}+\Omega_{3} \tau_{12}) \label{eq:gen2ndCor3}
 \\
&&\times
 {\rm Tr} \left(  \hat{E}^{+}
\left[  \left[\hat{E}^{-}   \left[\hat{E}^{-}   \hat{\rho}_{SS} \right](\Omega_{1}) 
  \right](\Omega_{2})  \hat{E}^{+} \right] (\Omega_{3})
 \right) \nonumber
\end{eqnarray}
for (iii) $\tau_2<\tau_1<\tau_3<\tau_4$. 
By inserting the Eqs.~(\ref{eq:gen2ndCor1})-(\ref{eq:gen2ndCor3}) into 
Eq.~(\ref{eq:2ndCorrelation}), we obtain a general expression

\begin{eqnarray} 
&& \left. \langle \mathcal{T}_+ \mathcal{T}_-
\hat{E}^{+}_F(t_1)    \hat{E}^{+}_F(t_2)  
\hat{E}^{-}_F(t_2)  \hat{E}^{-}_F(t_1) \rangle \right|_{t_1=t_2}  \nonumber \\
&&= 2 {\rm Re}\sum_{k={\rm i}}^{\rm iii} \sum_{\Omega_1,\Omega_2,\Omega_3} Z_{k} (\Omega_1,\Omega_2,\Omega_3)  \Theta_k (\Omega_1,\Omega_2,\Omega_3), \qquad
 \label{eq:2ndCorrelationGen}
\end{eqnarray}
where
\begin{eqnarray}
&& Z_{\rm i}=2^2 \iiiint_{({\rm i})} {\rm d}{\bf \tau}^4    f^\ast (\tau_1) f^\ast (\tau_2) f (\tau_3) f (\tau_4) \nonumber  \\
 && \qquad \times {\rm exp}\left( \Omega_{1} \tau_{14}+\Omega_{2}\tau_{43}+\Omega_{3} \tau_{32}\right) , 
 \label{eq:Zi}  \\
&& Z_{\rm ii}=2^2 \iiiint_{({\rm ii})} {\rm d}{\bf \tau}^4  f^\ast (\tau_1) f^\ast (\tau_2) f (\tau_3) f (\tau_4)  \nonumber  \\
&& \qquad \times  {\rm exp}\left(\Omega_{1} \tau_{41}+\Omega_{2}\tau_{13}+\Omega_{3} \tau_{32} \right) ,  \label{eq:Zii} \\
&& Z_{\rm iii}=2^2 \iiiint_{({\rm iii})} {\rm d}{\bf \tau}^4   f^\ast (\tau_1) f^\ast (\tau_2) f (\tau_3) f (\tau_4)  \nonumber  \\
&& \qquad \times  {\rm exp}\left(\Omega_{1} \tau_{43}+\Omega_{2}\tau_{31}+\Omega_{3} \tau_{12} \right) , 
\label{eq:Ziii}
\end{eqnarray}
and 
\begin{eqnarray}
&&\Theta_{\rm i}= {\rm Tr} \left(  \hat{E}^{+}
\left[ \hat{E}^{-} 
\left[ \hat{E}^{-} 
\left[ \hat{\rho}_{SS} \hat{E}^{+}  
\right](\Omega_{1})
\right](\Omega_{2})
\right](\Omega_{3}) \right),  \qquad  \\
&&\Theta_{\rm ii}= {\rm Tr} \left(  \hat{E}^{+}
\left[  \left[\hat{E}^{-}   \left[\hat{E}^{-}   \hat{\rho}_{SS} \right](\Omega_{1}) 
  \right](\Omega_{2})  \hat{E}^{+} \right] (\Omega_{3})  \right),  \qquad \\
&&\Theta_{\rm iii}= {\rm Tr} \left(  \hat{E}^{+}
\left[  \left[\hat{E}^{-}   \left[\hat{E}^{-}   \hat{\rho}_{SS} \right](\Omega_{1}) 
  \right](\Omega_{2})  \hat{E}^{+} \right] (\Omega_{3})
 \right) .  \qquad
\end{eqnarray}
With the general decomposed expression, the effect of the spectral filtering on the second-order correlation function enters only through $Z_k (\Omega_1,\Omega_2,\Omega_3)$ and $s(\Omega)$.  
Therefore, they can be regarded as response functions of the system in which the filter response is convolved.

We should mention here the case of short correlation time $\tau_c$ filters ($\tau_c$ is defined through $f(\tau \gg \tau_c)=0$), which should correspond to an unfiltered case.
If $\tau_c$ is much shorter than the time scale of system dynamics, we can put ${\rm exp}(\Omega_i \tau_j) = 1$ for $Z_k$ in Eqs.~(\ref{eq:Zi})-(\ref{eq:Ziii}) and for $s$ in Eq.~(\ref{eq:IntensityGenS}). 
In this case ($\tau_c \to 0$), $s$ and $Z_k$ for $k={\rm i}-{\rm iii}$ are independent on $\Omega$, $\Omega_1,\Omega_2$, and $\Omega_3$.
Then, by using $\sum_{\Omega_1,\Omega_2,\Omega_3}  \Theta_k =\langle 
\hat{E}^{+}  \hat{E}^{+}  
\hat{E}^{-}  \hat{E}^{-} \rangle$, $\sum_{\Omega}  q (\Omega) =\langle 
\hat{E}^{+}  \hat{E}^{-} \rangle$, we safely find that the expression for the normalized correlation function is reduced to be that of the unfiltered field,
\begin{eqnarray}
g^{(2)}_F (0)=\frac{\langle 
\hat{E}^{+}  \hat{E}^{+}  
\hat{E}^{-}  \hat{E}^{-} \rangle}{\langle 
\hat{E}^{+}  \hat{E}^{-} \rangle^2}.
\end{eqnarray}

{\it $s$ and $Z_k$ for Lorentzian/Gaussian/rectangular filters}---
As the typical examples, the above general expression is applied to three types of filters, Lorentzian, Gaussian, and rectangular filters, to obtain the explicit analytic forms for $s$ and $Z_k$ here.
In the calculation, we assume for simplicity that the time delay of the filter response, $\tau_d$ in Fig.~\ref{fig2} (a), is much larger than the correlation time, $\tau_c$, and in addition, the system is assumed to be in the steady state. Under this assumption, we will change the time variables from $\tau$ to $\tau +\tau_d$ and approximately change the lower limit of the time integration from $0$ to $-\tau_d \approx -\infty$.  With this change, the range of the integration for $s(\Omega)$ is replaced by $-\infty<\tau_1<\tau_2<\infty$ in Eq.~(\ref{eq:IntensityGenS}).
Similarly, for $Z_k$ in Eqs.~(\ref{eq:Zi})-(\ref{eq:Ziii}), the time range of the integration is replaced by $-\infty<\tau_j<\infty$ while the ordering among $\tau_1$, $\tau_2$, $\tau_3$, and $\tau_4$ are unchanged. 

The Lorentzian filter is the simplest example to perform the time integration to give $s$ $(= s^{L})$ and $Z_k$ $(= Z_k^L)$, since the correlation function of Lorentzian filter $f(\tau)= f_L(\tau)$ is an exponential~\cite{Nienhuis}, 
\begin{eqnarray}
f_L(\tau)=\lambda \theta(\tau) {\rm exp}\left( (-\lambda-i\omega_F)\tau \right), 
\end{eqnarray} 
where $\theta(x)$ is the Heaviside step function.
Inserting this and after straightforward integrations, we find that they are given by simple polynomial fractions,
\begin{eqnarray}
&& s^L(\Omega)=\frac{\lambda/2}{i\omega_F +\lambda -\Omega}, \\
&& Z^L_{\rm i} = \frac{\lambda}{\lambda-i \omega_F-\Omega_1} 
\frac{\lambda}{2\lambda-\Omega_2} 
\frac{\lambda}{3\lambda+i \omega_F-\Omega_3} , \quad \\
&& Z^L_{\rm ii} = \frac{\lambda}{\lambda +i \omega_F-\Omega_1} 
\frac{\lambda}{2\lambda-\Omega_2} 
\frac{\lambda}{3\lambda+i \omega_F-\Omega_3} , \quad \\
&& Z^L_{\rm iii} = \frac{\lambda}{\lambda+i \omega_F-\Omega_1} 
\frac{\lambda}{2\lambda+i 2 \omega_F-\Omega_2} 
\frac{\lambda}{3\lambda+i \omega_F-\Omega_3} . \nonumber \\ 
&& 
\end{eqnarray} 
For Lorentzian filters, the time integration for correlation functions gives the products of the transfer function, and therefore, analytic time integration upto the arbitrarily high orders is possible. Due to the simplicity, the photon statistics of filtered fields has been studied only for Lorentzian filters~\cite{Nienhuis, Valle}. However, as shown in the next section, when the time scale of the system dynamics is comparable to $\tau_c$, which we sometimes face in state-of-the-art quantum emitters, the best choice of the filter type is essential.
Therefore, the photon statistics of the field filtered by other types of filters should be necessary.
Here we just show the results for Gaussian and rectangular filters (but the method can be applied to arbitrary filter function).

For Gaussian filters, the correlation functions, $f(\tau)$ $(=f_G(\tau))$ in the time domain and $F(\omega )$ $(=F_G(\omega ))$ in frequency domain are, 
\begin{eqnarray}
f_G(\tau)&=&\frac{\lambda}{\sqrt{\pi}}{\rm exp}\left( -(\lambda \tau)^2 -i \omega_F \tau \right), \\
F_G(\omega)&=&\frac{1}{2 \pi}{\rm exp}\left( -\left (\frac{\omega-\omega_F}{2\lambda} \right)^2 \right), 
\end{eqnarray}
where the Fourier transform is defined by $F(\omega)\equiv (2 \pi)^{-1} \int f(\tau) {\rm exp}(i\omega \tau) {\rm d} \tau$. 
For this filter, $s(\Omega)$ $(=s^G(\Omega))$ is given by
\begin{eqnarray}
s^G(\Omega)&=& \frac{1}{2}{\rm exp}\left( y(\Omega)^2 \right) \left( 1+{\rm erf}\left( y(\Omega) \right)\right),
\end{eqnarray}
where $y(\Omega)\equiv (\Omega -i \omega_F)/(\sqrt{2} \lambda)$ and ${\rm erf}(x)$ is the Gauss error function. For the second-order correlation function, we obtained an analytic expression for $Z_k$ $(= Z_k^{G})$,
\begin{eqnarray}
Z^G_k &=&\frac{1}{\sqrt{\pi}}e^{A_k^2+B_k^2+C_k^2}\int_0^{\infty} e^{-(z-A_k)^2}\left(1- {\rm erf}(z+C_k)\right)  \nonumber \\
&& \times \left(
 {\rm erf}(z+B_k)- {\rm erf}(-z+B_k)
 \right){\rm d}z, \label{eq:ZkGauss}
\end{eqnarray}
whose coefficients, $A_k$, $B_k$, and $C_k$, are given in Table~\ref{table1}.
\begin{table}[tbp]
\caption{\label{table1} Coefficients for $Z^G_k$ of Gaussian filters, Eq.~(\ref{eq:ZkGauss}).}
\begin{ruledtabular}
\begin{tabular}{cccc}
$k$  & $A_k$ & $B_k$ &  $C_k$ \\
\hline
\hline
i &
  $\frac{\Omega_1-\Omega_2+\Omega_3}{2\lambda}$ &
  $\frac{-2i \omega_F-\Omega_1+\Omega_3}{2\lambda}$ & 
  $\frac{-\Omega_2}{2\lambda}$ \\
\hline
ii &
  $\frac{-2i \omega_F+\Omega_1-\Omega_2+\Omega_3}{2\lambda}$ &
  $\frac{-\Omega_1+\Omega_3}{2\lambda}$ & 
  $\frac{-\Omega_2}{2\lambda}$ \\
\hline
iii &
  $\frac{\Omega_1-\Omega_2+\Omega_3}{2\lambda}$ &
  $\frac{-\Omega_1+\Omega_3}{2\lambda}$ & 
  $\frac{2i \omega_F-\Omega_2}{2\lambda}$ \\
\end{tabular}
\end{ruledtabular}
\end{table}
\begin{table}[tbp]
\caption{\label{table2} Coefficients for $Z^r_k$ of rectangular filters, Eq.~(\ref{eq:ZkRect}).}
\begin{ruledtabular}
\begin{tabular}{cccc}
$k$  & $\alpha_k$ & $\beta_k$ &  $\gamma_k$ \\
\hline
\hline
i &
  $\frac{-\omega_F-i\Omega_3 }{\lambda}+i 0  $ &
  $\frac{-i \Omega_2 }{\lambda}+i 0$ & 
  $\frac{\omega_F+\lambda -i \Omega_1}{\lambda}+i0$ \\
\hline
ii &
 $\frac{-\omega_F-i\Omega_3 }{\lambda}+i 0  $ &
  $\frac{-i \Omega_2 }{\lambda}+i 0$ & 
  $\frac{-\omega_F+\lambda -i \Omega_1}{\lambda}+i0$ \\
\hline
iii &
$\frac{-\omega_F-i\Omega_3 }{\lambda}+i 0  $ &
  $\frac{-2\omega_F-i \Omega_2 }{\lambda}+i 0$ & 
  $\frac{-\omega_F+\lambda -i \Omega_1}{\lambda}+i0$  \\
\end{tabular}
\end{ruledtabular}
\end{table}

For rectangular filters, filter correlation functions, $f(\tau)$ $(=f_r(\tau))$ and $F(\omega )$ $(=F_r(\omega ))$ are given by   
\begin{eqnarray}
f_r(\tau) &=& {\rm exp}\left( -i \omega_F \tau \right)\frac{{\rm sin}(\lambda \tau)}{\pi \tau},  \label{eq:TimeFilterFuncRect} \\
F_r(\omega) &=& \frac{1}{2 \pi}\theta(\lambda-|\omega -\omega_F|).
\end{eqnarray}
For this filter, $s(\Omega)$ $(= s^r (\Omega))$ is found to be
\begin{eqnarray}
s^r(\Omega) &=&   \frac{1}{2\pi i}  
 \ln \left( \frac{\omega_F+\lambda+ i\Omega  -i 0}{\omega_F-\lambda+ i\Omega -i 0}\right),
\end{eqnarray}
where the infinitesimally small positive number, $0$, is introduced for the analytic continuation of the logarithmic function (which is essential in case ${\rm Re}(\Omega)=0$).
The analytic expression for $Z_k$ $( = Z_k^r)$ is also found by inserting Eq.~(\ref{eq:TimeFilterFuncRect}) into Eqs.~(\ref{eq:Zi})-(\ref{eq:Ziii}) and performing the integration,
\begin{eqnarray}
&& Z_k^r=\frac{i}{2\pi^3} \bigg( 
\big( \phi(\alpha_k^+,\beta_k^+;2)+\phi(\alpha_k^-,\beta_k^-;-2) \big)
\nonumber  \\
&& \qquad  
\times  \big( \ln(2-\gamma_k)-\ln (-\gamma_k) \big)  \nonumber
 \\
&& 
- \Phi(\alpha_k^+,\beta_k^+,\gamma_k;2)+\Phi(\alpha_k^-,\beta_k^-,\gamma_k-2;-2) 
\bigg) ,  \label{eq:ZkRect}
\end{eqnarray}
where $\alpha_k^\pm \equiv \alpha_k \pm 1$, $\beta_k^\pm \equiv \beta_k \pm 2$, and the coefficients $\alpha_k$, $\beta_k$, and $\gamma_k$ are given in Table~II. 
The functions, $\phi$ and $\Phi$, are defined with an analytically continued function of the $n$-th order polylogarithm, ${\rm Li}_n(z)=\sum_{m=1}^{\infty}z^m/m^2$, by
\begin{eqnarray}
&& \phi(a,b;z ) \equiv  -\ln (z-b)\ln(-a) \nonumber \\
&&\quad  +\ln(z-a) \ln \left(\frac{z-b}{a-b} \right) +{\rm Li}_2 \left( \frac{z-a}{b-a} \right),  \\
&& \Phi(a,b,c;z ) \equiv \int_0^z \frac{\phi(a,b;z )}{z-c}{\rm d}z. 
\end{eqnarray}
In the evaluation of $Z_k^r$, we have carefully performed the multiple complex integrations since the contours cross branch cuts of the logarithmic functions.

To summarize this section, a simple calculation method for photon statistics of filtered field, based on superoperator eigenvalue decomposition technique, was proposed and analytic expressions for $s$ and $Z_k$ are obtained for the three types of filters : Lorentzian, Gaussian, and rectangular filters (as typical examples).
For other types of filters, it will also be possible to find analytic expressions, although we will not go into further details here.
Validity of our method is confirmed numerically by the perfect agreements with the other method~\cite{Valle} for the case of Lorentzian filters, as shown in Sec.~\ref{sec: applications}.

\section{\label{sec: applications} Application to quantum dot single photon emitters}
Here, we take QDs as an example of efficient SP emitters and apply the proposed eigenvalue decomposition method to a simulation of the photon statistics of the emission field filtered by Lorentzian, Gaussian, and rectangular filters.

\subsection{\label{sec:QDemitter} QD SP emitters under incoherent pumping }

\begin{figure}[tb]
\begin{center}  
\includegraphics[width=8.5 cm]{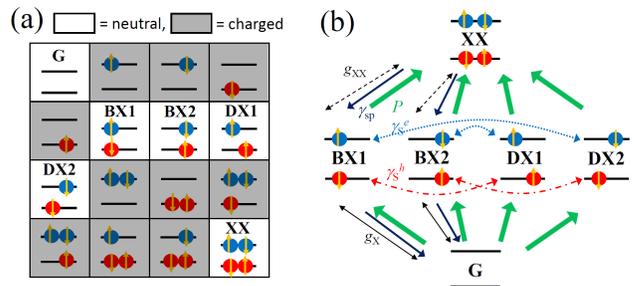} {} 
\end{center}
\vspace{-3mm} 
\caption{\label{fig3} A neutral QD model~\cite{Ritter, Yamaguchi, Kamide}. (a) Among 16 electronic states at the QD ground levels, 6 neutral states with upto two excitons (G, BX1, BX2, DX1, DX2, and XX) are taken into account.
(b) Incoherent pump ($P$) and decay processes ($\gamma_{sp}$, $\gamma_S^e$, $\gamma_S^h$).}
\end{figure}

The model of the QD emitter system is the same as that used in our previous paper~\cite{Kamide, Yamaguchi, Ritter}.
We consider the QD emitter states consisting of the electron-hole carriers as shown in Fig.~\ref{fig3} (a). 
Among sixteen carrier configurations occupying the lowest energy levels, six charge-neutral configurations are taken into account: 
an empty state, $| G \rangle $, two bright exciton (BX) states, $| BX1 \rangle ={\hat e}_\uparrow^\dagger {\hat h}_\downarrow^\dagger | G \rangle $ and $| BX2 \rangle ={\hat e}_\downarrow^\dagger {\hat h}_\uparrow^\dagger | G \rangle $, two dark exciton (DX) states, $| DX1 \rangle ={\hat e}_\uparrow^\dagger {\hat h}_\uparrow^\dagger | G \rangle $ and $| DX2 \rangle ={\hat e}_\downarrow^\dagger {\hat h}_\downarrow^\dagger | G \rangle $, and a biexciton state, $| XX \rangle ={\hat e}_\uparrow^\dagger {\hat e}_\downarrow^\dagger {\hat h}_\uparrow^\dagger {\hat h}_\downarrow^\dagger | G \rangle $, where ${\hat e}_\sigma$ and ${\hat h}_\sigma$ (${\hat e}_\sigma^\dagger $ and ${\hat h}_\sigma^\dagger $) are annihilation (creation) operators of electrons and holes with spin $\sigma=\uparrow, \downarrow$ in their respective lowest energy levels of the QD.
The Hamiltonian of the QD emitter is
\begin{eqnarray}
{\hat H}&=& \omega_X {\hat N}_{\rm tot} -\chi |XX \rangle \langle XX|,  \label{eq:HamiltonianQD}
\end{eqnarray}
where ${\hat N}_{\rm tot}=\sum_{\sigma=\uparrow,\downarrow} ({\hat e}_\sigma^\dagger {\hat e}_\sigma+{\hat h}_\sigma^\dagger {\hat h}_\sigma  )/2$ is the number of excitons, $\chi (= \omega_{X}- \omega_{XX})$ is the biexciton binding energy, and the fine structure splitting between the exciton states is neglected. 
The following incoherent decay processes as are considered shown in Fig.~\ref{fig3} (b):  the decay of the injected electron-hole pairs is dominated by the spontaneous emission (the rate $\gamma_{sp}$)~\cite{Johansen}, the excitons suffer dephasing (with the rate $\Gamma_{\rm ph}$), and the spin flip of electrons and holes (with the rates $\gamma_S^e$ and $\gamma_S^h$) results in the transitions between dark and bright exciton states with a rate $\gamma_S$ $(=\gamma_S^e+\gamma_S^h)$.

\begin{figure*}[tb] 
\begin{center}  
\includegraphics[width=15cm]{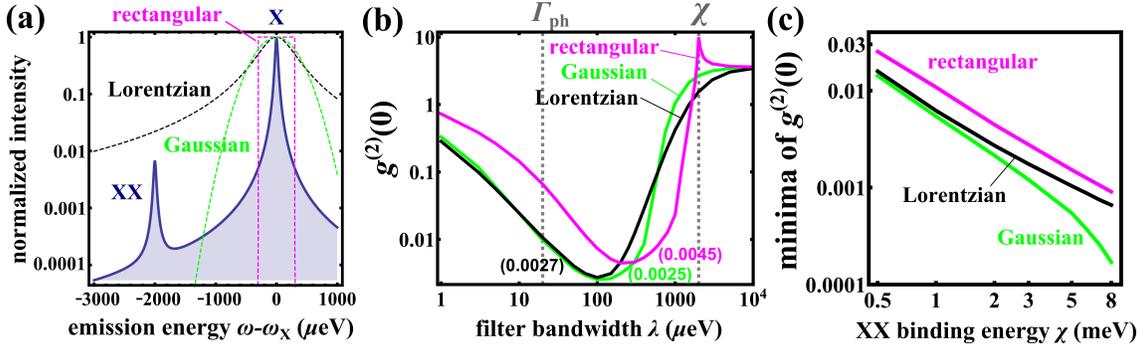} {} 
\end{center}
\vspace{-4mm} 
\caption{\label{fig4}  (a) Emission spectrum of a QD SP emitter (solid) for XX binding energy $\chi=2000$ $\mu$eV is shown with the normalized filter functions $|F(\omega)|^2$ (dashed) for Lorentzian (black), Gaussian (green), and rectangular (magenta) filters with $\lambda=300$ $\mu$eV.  (b, c) $g^{(2)}(0)$ of the QD emission after spectral filtering by the three types of filters, Lorentzian (black), Gaussian (green), and rectangular (magenta); (b) filter bandwidth ($\lambda$) dependence for XX binding energy $\chi=2000$ $\mu$eV, and (c) The $\chi$ dependence of the $g^{(2)}(0)$ at the optimal filter bandwidth $\lambda_{\rm opt}$ ($\approx 100$ $\mu$eV for Lorentzian filter in (b); $g^{(2)}(0)$ at $\lambda_{\rm opt}$ are also indicated in (b)). We set $(\gamma_{sp}, \ \Gamma_{\rm ph})=(0.67,20)$ $\mu$eV, small pump rate $P$ in the linear regime~\cite{Kamide}, and the spin flip time $\tau_S \equiv \gamma_S=10$ ns for all figures. }
\end{figure*}

In this neutral QD, it was shown that XX emission at $\omega=\omega_{X}-\chi$ is enhanced by incoherent XX excitation via DX states, and can strongly degrade the purity of SP emissions, especially in the case when the spin flip process is slower than the spontaneous emission ($\gamma_S<\gamma_{sp}$)~\cite{Kamide}. Therefore, if the exciton emission at $\omega=\omega_X$ is applied to a SP source, the XX emission must be effectively cut by using a spectral filter.
Here, for the calculation of the emission properties, we define the emission field operator for the BX recombination by ${\hat E}^- \equiv \sum_\sigma {\hat e}_\sigma {\hat h}_{-\sigma}=({\hat E}^+)^\dagger$, and the central frequency of the filter $\omega_F$ is set $\omega_F=\omega_X$.

In Fig.~\ref{fig4} (a), the emission spectrum is shown for a situation ($\chi=2$ meV, $\Gamma_{\rm ph}=20$ $\mu$eV, $1/\gamma_S=10$ ns, $1/\gamma_{sp}$=1 ns). 
In the same figure, the filter functions in the frequency domain $|F(\omega)|^2$ are also shown for Lorentzian, Gaussian, and rectangular filters with the bandwidth $\lambda=300$ $\mu$eV.
In the frequency domain, the Lorentzian filter has a long tail, Gaussian filter has a shorter tail, and rectangular filter has an ideally sharp cut.
Therefore, from the emission spectrum, the rectangular filter (with a bandwidth less than $\chi=2$ meV) may be expected to be the most effective filter, but we see in the following that the real situation is not so simple.

In Fig.~\ref{fig4} (b), we show the $g^{(2)}(0)$ obtained for the emission spectrally filtered by the three types of filters as a function of the bandwidth $\lambda$.
As expected, the $g^{(2)}(0)$ is reduced if the bandwidth is chosen as $\lambda<\chi$ for all filters. 
On the other hand, if the bandwidth is chosen too small, $g^{(2)}(0)$ increases as $\lambda$ decreases due to the increased time uncertainty ($\Delta t=\tau_c=1/\lambda$ in Fig.~\ref{fig2}) as mentioned above and in previous literature~\cite{Nienhuis}.
Therefore, by considering the two opposing effects, spectral suppression of the unwanted detection of XX emissions and increasing time uncertainty (decreasing time resolution) for too narrow filters, the existence of the optimal filter bandwidth $\lambda_{\rm opt}$ is expected.
As predicted from the above argument, we found the $g^{(2)}(0)$ shows the minima (=0.0027 for Lorentzian, 0.0025 for Gaussian, and 0.0048 for rectangular filters) in Fig.~\ref{fig4} (b) ($\Gamma_{\rm ph}<\lambda_{\rm opt}<\chi$ where $\Gamma_{\rm ph}$ gives the exciton linewidth).
Here, we note that our results for Lorentzian filter perfectly agree with those obtained by using the other method~\cite{Valle}, numerically showing the validity of our method.

The above findings, e.g. the existence of an optimal filter bandwidth at $\Gamma_{\rm ph}<\lambda_{\rm opt}<\chi$, seem to be trivial.
However, following findings are rather counter-intuitive; (I) the rectangular filter has the largest minimum value of $g^{(2)}(0)$ among the three filters although the rectangular filter ideally cuts the XX emission in the frequency domain, and (II) the optimal filter bandwidth $\lambda_{\rm opt}$ is much larger than the emission linewidth $\sim \Gamma_{\rm ph}$.
(I) also applies to a wide range of the XX binding energy (0.5 meV$<\chi<8$ meV) as seen in Fig.~\ref{fig4} (c), where the minima of $g^{(2)}(0)$ as a function of $\chi$ are shown for the three filters. 
Fig.~\ref{fig4} (c) shows that the Gaussian filter will be the best filter to purify the SP emission from this neutral QD system (after the optimization of the bandwidth).
(I) can be understood as the difference of the filter correlation function in the time domain, $f(\tau)$. 
The $f(\tau)(=f_r(\tau))$ in Eq.~(\ref{eq:TimeFilterFuncRect}) is the sinc function with the slow power-law decay at large $\tau$, different from the fast exponential decay for the other two filters. Therefore, the increase in the time uncertainty matters significantly if a rectangular filter is used.
In the case of the Gaussian filter, because the correlation function is Gaussian also in time domain, the long-time tail is strongly suppressed compared with the rectangular filter. Therefore, the lower value of $g^{(2)}(0)$ in Fig.~\ref{fig4} (c) with the Gaussian filter is reasonable.

\subsection{SP emitters under coherent pumping}
Our next example to study the filtering effect is a resonantly driven SP emitter. 
The resonantly scattered light by an emitter exhibits Mollow triplet emission spectrum~\cite{Mollow}, which can be applied to an indistinguishable SP source~\cite{Nienhuis, Muller, Ates, Buckley}, since the resonant excitation prevents emitters from suffering spectral diffusion and dark exciton effects, and by reducing dephasing processes. 

\begin{figure*}[tb] 
\begin{center}    
\includegraphics[width=14 cm]{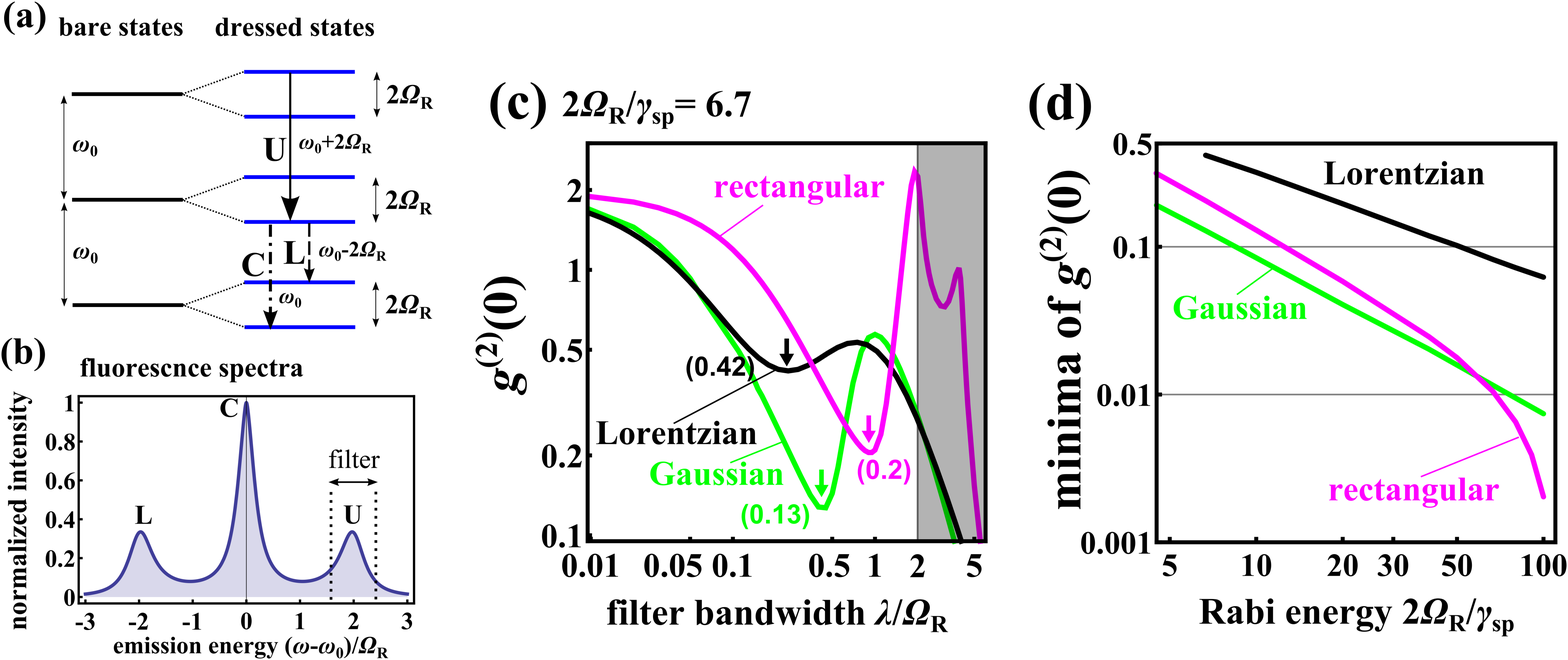} {} 
\end{center}
\vspace{-4mm} 
\caption{\label{fig5} Illustration of the single photon emission from coherently driven emitters (the Rabi frequency $\Omega_{\rm R}$ and $\omega_0 \equiv \omega_L=\omega_X$, laser frequency $\omega_L$, and the emitter transition frequency $\omega_X$). (a) the bare and dressed energy levels with the radiative transitions marked by arrows corresponding to the emission peaks in (b), and (b) the fluorescence spectrum showing Mollow triplets, a central peak (C) at $\omega=\omega_0$ and upper (U) and lower (L) side peaks. 
The $g^{(2)}(0)$ of the upper side peak emission (U) spectrally selected by three types of filters with $\omega_F=\omega_0+2 \Omega_{\rm R}$, Lorentzian (black), Gaussian (green), and rectangular (magenta). (c) $g^{(2)}(0)$ as a function of the normalized filter bandwidth $\lambda/\Omega_{\rm R}$. (d) $g^{(2)}(0)$ minimized in the range $0<\lambda<2 \Omega_{\rm R}$ as a function of the normalized rabi energy $\Omega_{\rm R}/\gamma_{\rm sp}$ (the values inside brackets and arrows indicate the minima in (c), the upper limit of the range is set in order to avoid the detection of the driving laser light in the shaded area, $\lambda>2 \Omega_{\rm R}$). 
We set the spontaneous emission $\gamma_{sp}=0.3 \Omega_{\rm R}$ and dephasing rate $\Gamma_{\rm ph}=0$ for (b) and (c), and the dephasing rate $\Gamma_{\rm ph}=0$ for all figures.}
\end{figure*}

The physics of the SP emission from the scattered light is illustrated in Fig.~\ref{fig5} (a).  
In the presence of a coherent laser field (frequency $\omega_L$) in resonance with the emitter ($\omega_X=\omega_L \equiv \omega_0$), the scattered light is known to exhibit the Mollow triplet~\cite{Carmichael} (Fig.~\ref{fig5} (b)) with a central (C) peak and two side peaks (L, U). The three peaks in the fluorescence spectrum correspond to the transitions (arrows) between the dressed states indicated by C (dash-dotted), L (dashed), U (solid), respectively in Fig.~\ref{fig5} (a).  From the illustration, the side peak (say the upper, U) is successively followed by emissions of the other side (L) or central (C) peaks. 
Therefore, successive two-photon emission within the same side peak is strongly suppressed if the splitting between the dressed states is larger than the line width ($2\Omega_{\rm R} \gg \gamma_{sp}$ if the linewidth is limited by the emitter lifetime).
This scheme to produce highly efficient and distinguishable single photons has been studied in
 recent years with QD emitters~\cite{Nienhuis, Muller, Ates, Buckley},
in which the spectral filtering of a side peak emission (say $\omega=\omega_0+2 \Omega_{\rm R}$) is essential.
In this scheme, the major cause of the contamination noise on the SP purity is the other two emission peaks ($\omega=\omega_0, \ \omega_0-2 \Omega_{\rm R} $) and the excitation laser itself ($\omega=\omega_0$).
The main physics can be described by a resonantly driven two-level system,
\begin{eqnarray}
\frac{\rm d}{{\rm d}t}{\hat \rho}
=i[{\hat \rho}, {\hat H}]+\gamma_{sp}{\mathcal L}_{{\hat \sigma}^-} {\hat \rho}+(\Gamma_{\rm ph}/2) {\mathcal L}_{{\hat \sigma}_z} {\hat \rho},
\end{eqnarray}
where ${\hat H}= \Omega_{\rm R} {\hat \sigma}^+ + \Omega_{\rm R}^\ast {\hat \sigma}^-$ in the rotating frame, and $\gamma_{sp}$ and $\Gamma_{\rm ph}$ are the spontaneous emission and dephasing rates.
Here we set the dephasing rate $\Gamma_{\rm ph} =0$ according to the experimental reports showing lifetime limited linewidth~\cite{Ates}.
For the calculation of the emission properties, we define the emission field operator by ${\hat E}^\pm \equiv {\hat \sigma}^\pm$, and the central frequency of the filter $\omega_F$ is set $\omega_F=\omega_0+2 \Omega_{\rm R}$.

Figure~\ref{fig5} (c), shows the simulated $g^{(2)}(0)$ of the upper side peak emissions spectrally filtered by the three types of filters, Lorentzian (black), Gaussian (green), rectangular (magenta) filters. 
The $g^{(2)}(0)$ is plotted as a function of the filter bandwidth $\lambda$, which shows a minimum in the regime $0<\lambda <2 \Omega_{\rm R}$ reflecting the physics of SP generation mentioned above;
If $\lambda <2 \Omega_{\rm R}$, $g^{(2)}(0)$ decreases as $\lambda$ decreases since the spectral selection for the side peak becomes effective. 
On the other hand, for $\lambda$ much less than the spontaneous emission rate $\gamma_{sp}$ or the linewidth ($\gamma_{sp} =0.3 \Omega_{\rm R}$ in Fig.~\ref{fig5} (c)), $g^{(2)}(0)$ increases as $\lambda$ decreases due to the degraded time resolution.
The $g^{(2)}(0)$ at the minima depends on the types of the spectral filters, being similar to Fig.~\ref{fig4} (b), the result for an incoherely pumped QD.
In this case, however, the Gaussian filter gives the smallest $g^{(2)}(0)$ i.e. the most pure SP emission, and the Lorentzian filter gives the worst purity for $\gamma_{sp} =0.3 \Omega_{\rm R}$  ($2 \Omega_{\rm R}/\gamma_{sp}=6.7$) in Fig.~\ref{fig5} (c).
Here, we again note that  our results for Lorentzian filter perfectly agree with those obtained by using the other method~\cite{Valle}, numerically showing the validity of our method. 
 
In Fig.~\ref{fig5} (d), the $g^{(2)}(0)$ at the minima is shown as a function of the ratio between the Rabi frequency over the linewidth, $2 \Omega_{\rm R}/\gamma_{sp}$.
It is clearly found that $g^{(2)}(0)$ at the minima decreases (i.e. the SP purity increases) as $2\Omega_{\rm R}/\gamma_{sp}$ increases. 
This is because the contamination source i.e. the other emission peaks, C and L, become spectrally separated and suppressed well by the filters for larger splitting, $2\Omega_{\rm R}/\gamma_{sp}$.
An interesting feature is that the Lorentzian filter gives $g^{(2)}(0)$ larger than others i.e. the performance to obtain high SP purity is the lowest among the three types, while the rectangular filter was the worst choice in the case of incoherent excitation in Fig.~\ref{fig4} (c).
Moreover, an interesting result is that the most efficient filter type to give the highest SP purity depends on the pump parameter $2\Omega_{\rm R}/\gamma_{sp}$ in Fig.~\ref{fig5} (d).
For the Rabi splitting not too large $2\Omega_{\rm R}/\gamma_{sp}<60$, the Gaussian filter with the smallest $g^{(2)}(0)$ is the best choice among the three. For the strong Rabi field $2\Omega_{\rm R}/\gamma_{sp}>60$, the rectangular filter with the smallest $g^{(2)}(0)$ is the best filter type. We should note here that the latter case especially is quite different from the result of incoherent excitation in Fig.~\ref{fig4} (c).

A simple explanation for this nontrivial matching between the filter and emitter remains elusive. 
In general, the matching between filter types and emitters will depend on the non-universal details of the pumping and emission dynamics.
Therefore, the numerical simulation under given conditions and the direct comparison between different filters are necessary to tailor and optimize the quantum emissions.

\section{\label{sec: summary}Summary} 
We proposed a calculation method, based on the superoperator eigenvalue decomposition technique, for photon statistics of spectrally filtered fields with various types of filters. This method can give exact results when the emission dynamics is given by quantum master equations, which can be applied to a wide variety of quantum emitters, and solvable with the eigenvalue approach (matrix diagonalization).
Also, it is possible to treat a wide variety of filter function if analytic expressions for the convolution functions, $s$ and $Z_k$ in Eq.~(\ref{eq:IntensityGenS}) and Eqs.~(\ref{eq:Zi})-(\ref{eq:Ziii}), are obtained.

As typical examples, focusing on three filter types, Lorentzian, Gaussian, and rectangular filters, we applied this method to QD single photon (SP) emitters. 
With the simulation for two cases, under incoherent excitations and under coherent and resonant excitations, we found condition-dependent non-universal matching between filter-types and emitters in order to have the highest SP purity.

An interesting issue remaining will be extending this method to simulations for periodic and short-pulsed pumping, which will allow us to study the effect of the spectral filtering in case of short-pulse excitations aiming at more realistic operations of the QD SP source~\cite{Kamide}.

\acknowledgements
We thank Y. Ota, M. Holmes, T. Rae, S. Kako, T. Miiyazawa, M. Yamaguchi, and T. Horikiri for useful comments and discussions.
This work is supported by the Project for Developing Innovation Systems of MEXT. This work is also supported by JSPS KAKENHI Grant Number 15K20931.

\end{document}